\documentclass[runningheads]{svmult} 
\usepackage{makeidx}   
\usepackage{graphicx}  
\usepackage{subeqnar}  
\usepackage{multicol}  
\usepackage{cropmark} 
\usepackage{physprbb}  
\newcommand{\nue}{\nu_e} 
\newcommand{\num}{\nu_\mu} 
\newcommand{\nut}{\nu_\tau} 
\newcommand{\nus}{\nu_s}

\newcommand{\rar}{\rightarrow}

\begin{document} 
\title*{Big Bang Nucleosynthesis and neutrinos} 
\toctitle{Big Bang Nucleosynthesis and neutrinos} 
%
%
%
\author{F.L.~Villante\inst{1} and  
A.D.~Dolgov\inst{1} \inst{2}} 
\authorrunning{F.L.~Villante et al.} 
%
%
\institute{Dipartimento di Fisica and Sezione INFN di Ferrara,  
Via del Paradiso 12, I-44100 Ferrara, Italy  
\and ITEP, Bol. Cheremushkinskaya 25, 117218, Moscow, Russia} 
 
\maketitle              
 
\begin{abstract} 
We present a brief review of Big Bang Nucleosynthesis (BBN).  
We discuss theoretical and observational uncertainties  
in deuterium and helium-4 primordial abundances and their  
implications for the determination of important cosmological parameters.  
We present, moreover, some recent results on active-sterile neutrino 
oscillations in the early universe and on their effects on BBN. 
\end{abstract}

\section{Introduction} 
Big Bang Nucleosynthesis (BBN), as well known, is one  
of the solid pillars of the standard  
cosmological model. The theory predicts that relevant 
abundances of light elements, namely $^{2}{\rm H}$, $^{3}{\rm He}$, 
$^{4}{\rm He}$ and $^{7}{\rm Li}$, have been produced during the first 
minutes of the evolution of the Universe. The predictions span about 
9 orders of magnitude and are in reasonable agreement with observations. 
Theoretical calculations are well defined and 
very precise. The largest uncertainty arises from the values of 
cross-sections of the relevant nuclear reactions.  
Theoretical accuracy is at the 
level of 0.2\% for $^{4}{\rm He}$, 5\% for $^{2}{\rm H}$ and $^{3}{\rm He}$  
and 15\% for $^{7}{\rm Li}$.  
However, comparison of theoretical results 
with observational data is not straightforward because the data 
are subject to poorly known evolutionary effects and systematic errors. 
Still, even with these uncertainties, BBN permits to constraint important 
cosmological parameters and to eliminate many modifications of the standard 
model, thus allowing to derive restrictions on the properties  
of elementary particles and, in particular, of neutrinos. 
 
In this paper, we briefly review the physics of BBN. 
In sect.~\ref{bbn-phys} we introduce the essential parameters  
and inputs. 
In sect.~\ref{obs} we summarize the present situation of  
observational data. 
In sect.~\ref{param} we discuss the determination of  
cosmological parameters.  
The last section is dedicated to 
BBN bounds on non-standard neutrinos and, specifically, to BBN and  
neutrino oscillations\footnote{In this paper, due to space limitation, 
we will consider only selected topics. For a complete review of the BBN 
bounds on neutrinos see ref.\cite{dolgov-rev}}.

\begin{figure}[t] 
\begin{center} 
\includegraphics[width=.7\textwidth]{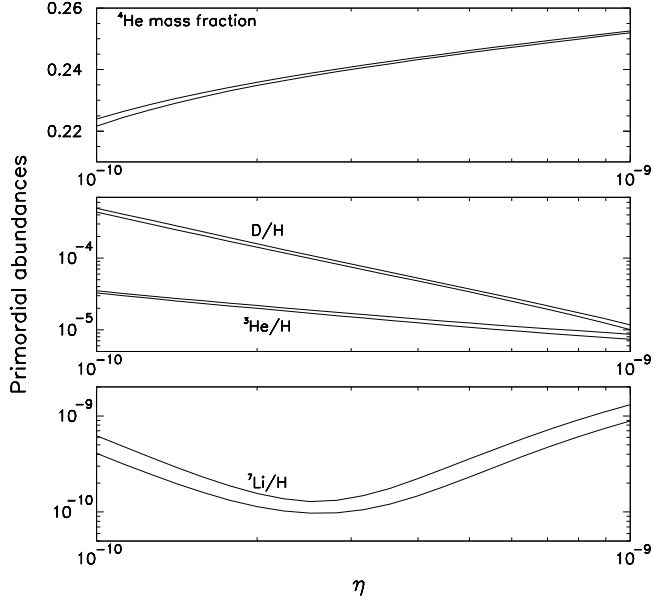} 
\end{center} 
\caption[]{Primordial light element abundances as predicted by standard BBN.
The widths of the bands correspond to theoretical uncertainties.} 
\label{f1} 
\end{figure} 
 
\section{The Physics of BBN} 
\label{bbn-phys} 
 
To understand primordial nucleosynthesis, we must follow in 
detail the histories of nucleons in the early universe. This 
is usually done by using numerical codes (among which 
the Fortran code by Wagoner \cite{wag}, updated by Kawano \cite{kaw} 
has become a standard tool). However, the main features of Big-Bang 
nucleosynthesis can be obtained by simple analytic arguments described  
in this section  
(see \cite{wei,kolb,sarkar-rev,stei-rev,sar-proc,stei-proc}  
for details).  
 
Primordial nucleosynthesis occurs at temperatures $T\le1$ MeV, which are  
small with respect to nucleon masses. At these temperatures, the number 
of nucleons is simply equal to the initial baryon asymmetry  
of the universe. It is useful to describe this quantity in terms of 
the {\it present} baryon to photon ratio: 
\begin{equation} 
\eta\equiv(N_{B}-N_{\overline{B}})/N_{\gamma} ~. 
\end{equation} 
The parameter $\eta$ is simply related to the baryon density of 
the universe, being $\Omega_{\rm B} h^{2} = 3.7\cdot 10^{7} \,\eta$. 
 
The neutron to proton ratio is controlled by the weak processes   
\begin{eqnarray} 
\nonumber 
n+e^{+} &\leftrightarrow& p +\overline{\nu}_{e}\\ 
\nonumber 
n+\nu_{e} &\leftrightarrow& p + e^{-}\\ 
n &\rightarrow& p + e^{-} +\overline{\nu}_{e} 
\label{np-weak} 
\end{eqnarray} 
which interconvert neutron and proton. When the temperature  
$T$ of the universe is about 1 MeV, these reactions 
are fast enough to maintain neutron and proton in  
chemical equilibrium. The neutron abundance is thus given by: 
\begin{equation} 
X_{n}\equiv\frac{n_{n}}{n_{n}+n_{p}} =  
\frac{1} 
{1+\exp(\Delta m/ T + \xi_{e})} 
\label{np-eq} 
\end{equation} 
where $\Delta m = 1.29$ MeV is the neutron-proton 
mass difference and $\xi_{e}=\mu_{e}/T$ is the dimensionless chemical  
potential of electron neutrinos (in standard BBN $\xi_{e}$ is assumed  
to be negligible).  
 
When the temperature $T$ drops below 
$T_{\rm f}=0.6-0.7$ MeV, the neutron-proton 
inter-conversion rate,  
$\Gamma_{\rm W}\sim G_{\rm F}^{2} T^{5}$,   
becomes smaller than the expansion rate 
the universe, $H\sim\sqrt{g_{*}G_{\rm N}}T^{2}$, 
where $g_{*}$ counts the total number of 
relativistic degrees of freedom of the early universe. 
The deviation of $g_{*}$ from the standard value, $g_{*}=10.75$,  
is usually described in terms of an {\it equivalent} number of massless 
neutrinos $N_{\nu}\neq3$ according to: 
\begin{equation} 
g_{*} = 10.75 + \frac{7}{4}(N_{\nu}-3) ~. 
\end{equation} 
 
For temperatures $T\le T_{\rm f}$, chemical equilibrium can  
no longer be maintained. The neutron abundance 
$X_{n}$ evolves only due to neutron decay, according  
to $X_{n}=X_{n}(T_{\rm f})\exp(-t/\tau_{n})$, 
where $\tau_{n}$ is the neutron lifetime.  
One should note that the ``freeze-out'' temperature  
$T_{\rm f}$ 
scales as $T_{\rm f}\propto g_{*}^{1/6}$ and thus is sensitive to the  
particle content of the early universe.  
The larger is $g_{*}$ (or equivalently $N_{\nu}$), the earlier is 
the freeze-out of the neutron abundance, at an higher value, and hence, 
the larger is the $^{4}{\rm He}$ abundance produced in BBN. 
 
When the temperature of the universe is equal to 
$T_{N}\simeq 0.06-0.07$~MeV neutrons and protons start to react each  
other to build up light nuclei.  
The exact value of $T_{N}$ depends on  
the baryon to photon ratio $\eta$. 
Only two body reaction are indeed important in BBN, 
such as $p(n,\gamma)^{2}{\rm H}$,  
$^{2}{\rm H}(p,\gamma)^{3}{\rm He}$, 
$^{3}{\rm He}(d,p)^{4}{\rm He}$, etc. 
(see \cite{smith}).  
Deuterium must be produced in appreciable quantity before  
the other reactions can proceed at all. 
However, due to the large number of photons per baryon,  
photodissociation of deuterium is not suppressed until the temperature  
decreases well below the deuterium binding energy $B_{d}=2.2$MeV. 
Following \cite{kolb}, one can see that the temperature $T_{N}$ below  
which deuterium production is favoured scales as  
$T_{N}\sim B_{d}/(15-\ln\eta)$. 
 
Once deuterium is formed, nucleosynthesis begins and light nuclei 
are produced rapidly. Essentially all available nucleons are quickly 
bound into $^{4}{\rm He}$, which is the most tightly bound  
light nuclear species. In addition to $^{4}{\rm He}$,  
substantial amounts of $^{2}{\rm H}$,  
$^{3}{\rm He}$ and $^{4}{\rm He}$ are produced. 
No heavy elements ($A>8$) are produced, due both to the fact that  
Coulomb-barrier suppression is very significant and to the absence of 
stable isotopes with $A=5$ and $A=8$. 
In fig.1 we show the light element abundances produced during BBN, 
as calculated by using the Kawano code \cite{kaw}, for  
$\eta$ between $10^{-10}$ and $10^{-9}$. 
The calculation of $^{4}$He abundance includes  
small corrections due to radiative processes at zero and 
finite temperature, non-equilibrium neutrino heating during $e^{\pm}$ 
annihilation, and finite nucleon mass effects \cite{lopez,miele}.   
 
Theoretical predictions are affected by uncertainties at the level 
of 0.2\% for $^{4}{\rm He}$, 5\% for $^{2}{\rm H}$ and $^{3}{\rm He}$  
and 15\% for $^{7}{\rm Li}$. These uncertainties are due to  
uncertainties in the weak rates (which are ``normalized'' to the  
measured neutron lifetime $\tau_{n}=885.7\pm 0.8~{\rm s}$, see 
\cite{sarkar-rev} for details)  
and in the values of the relevant nuclear reaction 
rates. They have been estimated by montecarlo or semi-analytical  
methods \cite{smith,fiorentini}.  
Recently the nuclear data have been 
re-analyzed, leading to improved precision in the abundance predictions  
\cite{olive,burles,miele2}.

\section{Observational Data} 
\label{obs} 
 
The abundances of light elements synthesized in the Big Bang have been 
subsequently modified through chemical evolution of the astrophysical  
environments where they are measured. The observational strategy is to 
identify sites which have undergone as little chemical processing as  
possible and rely on empirical methods to infer the primordial  
abundances. For example, measurements of deuterium are made in 
in quasar absorption line systems (QAS) at high redshift; if there  
is a ``ceiling'' to the abundance in different QAS then it can be  
assumed to be the primordial value. The $^4{\rm He}$ abundance  
is measured H II regions in blue compact galaxies (BCGs) which have  
undergone very little star formation. Its primordial value is 
inferred either by using the associated nitrogen or oxygen abundance to 
track the stellar production of helium, or by simply observing the 
most metal-poor objects.  
Closer to home, the observed uniform abundance of $^{7}{\rm Li}$ 
in the hottest and most metal poor Pop II stars in our Galaxy is 
believed to reflect its primordial value.  
(We do not consider $^{3}{\rm He}$ whose post-BBN evolution is more 
complex.)   
 
As observational methods have become more sophisticated, 
the situation has become more, instead of less, complex. 
Relevant discrepancies, of a systematic nature, have emerged between 
different observers. In the following, we give a brief summary  
of the present situation for deuterium and helium-4  
(looking from outside by a non-expert). We refer to \cite{stei-rev}  
for a more complete and up-to-date discussion.

\vspace{0.4cm} 
\par\noindent 
{\it Deuterium}\\ 
Post-BBN evolution of deuterium (D) is simple. Deuterium is burnt in stars 
producing $^{3}{\rm He}$. Any deuterium measurement provides thus a lower  
limit for the primordial D abundance and an upper limit for baryon density 
of the universe.   
 
In recent years, measurements of deuterium have been made 
in quasar absorption line systems (QAS) at high redshift. 
These systems are presumably not contaminated by stellar  
processes and thus the observed deuterium should be close to 
the primordial one. 
Since deuterium isotope shift corresponds to velocity  
of only $(-82)~{\rm km/sec}$, 
clearly QAS with simple velocity structure are needed for reliable 
determinations. Moreover, ionization corrections, possible ``interlooper''  
etc. further increase systematic uncertainties. 
 
In tab.~\ref{deuterium} we give the results of recent deuterium 
determinations  in QAS \cite{d1,d2,d3,d4,d7,d5,d6}.  
An estimate of primordial deuterium abundance 
can be obtained from the weighted mean of data in tab.~\ref{deuterium}. 
It should be noted, however, that the dispersion among the  
different determinations is not consistent with 
errors in the single measurements (see ~\cite{d7} for 
detailed discussion).  
We will use, in the following, the value  
${\rm D/H}=2.78^{+0.44}_{-0.38}\times 10^{-5}$  
given in \cite{d7}, which is the weighted mean of the log~D/H values  
given by 
\cite{d1,d2,d3,d4,d7}. The quoted error is the 1$\sigma$ error in the mean,  
given by the standard deviation of the five log D/H values divided by  
$\sqrt{5}$. This error is used instead of the usual error in the  
weighted mean, in order to take into account  
the ``anomalous'' dispersion of deuterium data.

\begin{table} 
\caption{Deuterium abundance in quasar absorption line  
systems at high red-shift (see \cite{d7} for details).} 
\begin{center} 
\renewcommand{\arraystretch}{1.4} 
\setlength\tabcolsep{5pt} 
\begin{tabular}{lllllll} 
\hline\noalign{\smallskip} 
$z$ & 2.504 \cite{d1} & 3.572 \cite{d2} & 2.536 \cite{d3}  
& 2.076 \cite{d4}  
& 2.526 \cite{d7} & $3.025^{\mathrm a}$ \cite{d5} \\ 
$10^{5}({\rm D/H})$  
& $3.98^{+0.59}_{-0.67}$ & $3.25\pm0.3$ & $2.54\pm0.23$  
&$1.65\pm0.35$ 
& $2.42^{+0.35}_{-0.25}$ & $3.75\pm0.25$\\ 
\hline 
\end{tabular} 
\end{center} 
$^{\mathrm a}$ This system was first analyzed by \cite{d6} with the result  
$D/H = (2.24\pm0.67)\times 10^{-5}$. The quoted value is from \cite{d5}. 
\label{deuterium} 
\end{table}

\vspace{0.4cm} 
\par\noindent 
{\it Helium-4}\\ 
As a result of stellar processing, $^{4}\rm{He}$ is produced, 
increasing its abundance above the primordial value, 
together with ``metals'', such as nitrogen, oxygen and 
other elements heavier than $^{4}\rm{He}$, which are not produced in BBN. 
The observed  $^{4}\rm{He}$ abundance provides thus an upper bound to 
the primordial one, $Y_{\rm p}$.  
 
Helium observations are done in H II regions in blue compact  
galaxies (BCGs) which have undergone very 
little star formation (at present $\sim 100$ H II regions have been  
studied for their helium content). 
In order to infer the primordial value $Y_{\rm p}$, one extrapolates  
to zero metallicity ($Z=0$) the observed relation between  
helium ($Y$) and metals ($Z$). 
This is usually done using by nitrogen ($N$) or oxygen ($O$) as metallicity  
tracers. Alternatively, one can simply average helium abundances in  
most metal poor objects. 
 
The present situation is that independent determinations of  
$Y_{\rm p}$ have a statistical errors at the level of $1-2\%$ but  
differ among each others by about $\sim 5\%$. In particular, by 
using independent data sets, Olive and Steigman~\cite{olive-he1} and  
Olive et al.~\cite{olive-he2} have obtained $Y_{\rm p} = 0.234 \pm 0.003$,  
while Izotov et al.~\cite{izotov-he1} and Izotov 
and Thuan~\cite{izotov-he2} have found $Y_{\rm p} = 0.244 \pm 0.002$. 
 
The discrepancy between different determinations is possibly  
related to different description of the complex physical processes  
acting in H II regions. Several sources of systematic uncertainties 
may, in fact, affect the helium determination at a level comparable 
or larger than the reported statistical errors, like e.g.  
the evaluation of the ionization correction factor (which is related to  
how much neutral helium is in the object under scrutiny), 
of the temperature correction factor, of underlying  
stellar absorption, etc. (see \cite{luridiana} for a review). 
  
As discussed in the next section, it is extremely important to have  
a better determination of the $^{4}{\rm He}$ primordial abundances  
and a reliable estimate of the total (statistical + systematic)  
associated error. For our estimates,  
we will use the central values for $Y_{\rm p}$ reported  
above ($Y_{\rm p}=0.234$ and $Y_{\rm p}=0.244$),  
and the error estimate $\Delta Y_{\rm p} =0.005$, which is  
obtained e.g. in \cite{stei-rev,sar-proc} from the dispersion 
of the various $Y_{\rm p}$ determinations.

\section{Cosmological parameters from BBN} 
\label{param} 
 
The deuterium abundance ${\rm D/H}=2.78^{+0.44}_{-0.38}\times 10^{-5}$ 
can be used to determine the baryon density of the universe.  
As discussed in \cite{d7}, the quoted value corresponds 
to $\eta=5.9\pm0.5\times10^{-10}$ (in standard BBN)  
or, equivalently, to  
$\Omega_{\rm B}h^{2} = 0.0214 \pm 0.0020$.  
The error budget is dominated by the observational uncertainties  
which are about a factor 3 larger than uncertainties in  
theoretical prediction. 
 
 The obtained value for $\Omega_{\rm B}h^{2}$ has to be compared with 
independent determination of the baryon density of the universe. 
In particular with the result $\Omega_{\rm B}h^{2}=0.0224\pm0.0009$  
given in \cite{wmap} which is obtained from a combined fit to the cosmic  
microwave background (CMB) and large scale structure (LSS) data.  
The agreement of these two independent determinations is extremely  
important because they rely on completely different  
physical phenomena (which occurred at different time during the 
evolution of the universe). 
We note that CMB (and LSS) are presently more accurate than  
BBN in determining the baryon density of the universe.  
 
 The value of $\eta$ deduced from deuterium can be used, in standard BBN, 
to predict the abundance of the other elements and to compare with 
observations. Following \cite{d7}, one obtains $Y_{\rm p}=0.2476 \pm 0.0010$,  
$^{3}{\rm He}/{\rm H}=1.04\pm0.06 \times 10^{-5}$ and  
$^{7}{\rm Li}/{\rm H}=4.5\pm0.9 \times 10^{-10}$. 
It is evident that there is tension between the quoted values  
and the observational results. The ``predicted''  abundance 
for $^{4}{\rm He}$ is higher than the ``high'' helium value of 
Izotov et al. \cite{izotov-he1,izotov-he2}. Moreover, the ``predicted'' 
$^{7}{\rm Li}$ abundance is a factor 2-3 larger with respect to  
the present observational results \cite{lithium1}. 
The origin of these differences has to be clarified. They could be due to  
systematic errors in the measurements or to evolutionary effects  
(e.g. $^{7}{\rm Li}$ depletion) or they could be a real  
indication for non-standard effects in BBN. 
 
In particular, the present D and $^{4}{\rm He}$ data seems to 
favour an equivalent number of neutrino families $N_{\nu}\le3$.  
In order to understand the present situation, it 
is useful to combine the deuterium value  
${\rm D/H}=2.78^{+0.44}_{-0.38}\times10^{-5}$,  
with the ``low'' helium abundance, $Y_{\rm p}=0.234\pm0.005$,  
or with the ``high'' helium abundance, $Y_{\rm p}=0.244\pm0.005$.  
The error $\Delta Y_{\rm p}=0.005$ is the ``estimated''  
systematic error in $^{4}{\rm He}$ measurements (see above). 
If we fit these data in the plane  $(\eta,N_{\nu})$  
following \cite{sarkar-noi}, we obtain the bound  
$N_{\nu} = 2.3\pm 0.5$ ($1\sigma$) in the first case,  
and $N_{\nu} = 2.8 \pm 0.5$ ($1\sigma$) in the second. 
In both cases, the central values are below three, even 
if the errors  are large enough to allow for the  
standard value $N_{\nu}=3$. 
  
The described results clearly indicate that 
a large number of effective neutrinos is 
disfavoured.  
One can conclude, in principle, that an upper bound on the number  
of {\it extra} neutrinos, $\delta N_{\nu}\equiv N_{\nu}-3$, is 
$\delta N_{\nu}\le0.3$. 
It is clear, however, that the situation is quite delicate. 
The error $\Delta N_{\nu}=0.5$ is completely dominated by 
systematic error in $^{4}{\rm He}$ measurements. 
For this reason, we believe that, at present stage, a more 
safe upper bound on the number of extra neutrinos is $\delta N_{\nu}\le 1$. 
Hopefully in the near future we will be able to derive a stronger 
limit.

Other physical parameters which can be bounded by BBN are  
the chemical potentials of different neutrino species,  
$\mu_{a}$ where $a=e,\mu,\tau$. 
The possible role of neutrino degeneracy in BBN was noted already in 
\cite{chem1} and then discussed in a number of papers. A non vanishing 
chemical potential for $\nu_{e}$, $\nu_{\mu}$ or $\nu_{\tau}$ 
would increase the neutrino contribution to the energy density 
and can be described as an increase in $N_{\nu}$. 
An additional (and dominant) effect exists for electron neutrinos  
which directly participate to n-p interconversion reactions. 
A non vanishing $\mu_{e}$ would shift the equilibrium between neutrons 
and protons, see eq.~(\ref{np-eq}), with large effects on light elements 
production. 
  
Several analysis have been made of the BBN limits on neutrino chemical 
potentials. A recent analysis which include also CMB data \cite{chem2} 
concludes: 
\begin{eqnarray} 
-0.01 < &\xi_{e}& < 0.2 \\ 
&|\xi_{\mu,\tau}|&<2,6 
\end{eqnarray} 
where $\xi_{a}= \mu_{a}/T$ are the dimensionless chemical potentials. 
For further implications and for a discussion of the case in which  
both $N_{\nu}$ and $\xi_{e}$ are free to vary see \cite{chem3}.

\section{BBN and neutrino oscillations} 
 
Effects of neutrino oscillations on BBN are much different if 
only active neutrinos are mixed, if only one active and one 
sterile neutrino are mixed or if we consider the more ``complete'' 
case of mixing between three active and one sterile neutrino.  
 
\subsection{Mixing between active neutrinos} 
 
If neutrinos are in thermal equilibrium with vanishing 
chemical potentials, mixing between active neutrinos does not  
introduce any deviation from standard BBN results.  
 The situation is more interesting if neutrinos are degenerate.  
In particular, it was shown recently \cite{chem+osc} that,  
for the mixing parameters which explain the solar neutrino  
problem \cite{deholanda}
($\delta m^2_{\rm sol}=7.3\cdot 10^{-5}\;{\rm eV}^2 
$ and $\tan^{2}\theta_{\rm sol}=0.4$)  
and the atmospheric neutrino anomaly \cite{nu-atmo}
($\delta m^2_{\rm atmo}=2.5\cdot 10^{-3}\;{\rm eV}^2$ and  
$\tan^{2}\theta_{\rm atmo}\approx 1$),  
asymmetries in the muonic and/or tauonic neutrino sectors would  
produce, through oscillations, an asymmetry into the electronic neutrino  
sector. This means that, in presence of oscillations, 
the restrictive bounds on the chemical potential of electron neutrinos 
applies to all neutrino flavours. It is thus possible to obtain the  
restrictive bound: 
\begin{equation} 
|\xi_{a}| < 0.1 ~, 
\end{equation} 
valid for any flavour \cite{chem+osc}.

\subsection{Mixing between one active and one sterile neutrino}

There are three possible effects on BBN created by mixing 
between active and sterile neutrinos. First is the production of additional 
neutrino species in the primeval plasma, leading to $N_{\nu}>3$.  
The second effect is a depletion of the number density of electronic neutrinos 
which results in a higher neutron freezing temperature. Both these 
effects lead to a larger neutron-to-proton ratio and to  
more abundant production of primordial deuterium and helium-4 (for  
the details see e.g. review~\cite{dolgov-rev}). 
If mixing between active neutrinos is absent the second  
effect would manifest itself only in the case of $(\nue-\nus)$-mixing, 
if we neglect relatively weak depopulation of $\nue$ through the annihilation 
$\bar\nue \nue \rar \bar\nu_{\mu,\tau} \nu_{\mu,\tau}$.  
 
The third effect is a generation of large lepton asymmetry due to  
oscillations between active and sterile species~\cite{res-rise}. 
However, this effect takes place only for very weak mixing, much 
smaller than the experimental bound and is not discussed
in this paper. 
 
The problem of active-sterile neutrino oscillation is quite complex  
and has been discussed in many papers starting from 1990  
(a large list of references can be found in ref.~\cite{dolgov-rev}).  
The problem was recently re-considered in \cite{dolgovil} both 
analytically and by solution of the complete system of integro-differential  
kinetic equations. 
Earlier derived bounds 
have been re-analyzed and significantly  
different results have been found in the resonance case. 
 
\begin{figure}[t] 
\begin{center} 
\includegraphics[width=.8\textwidth]{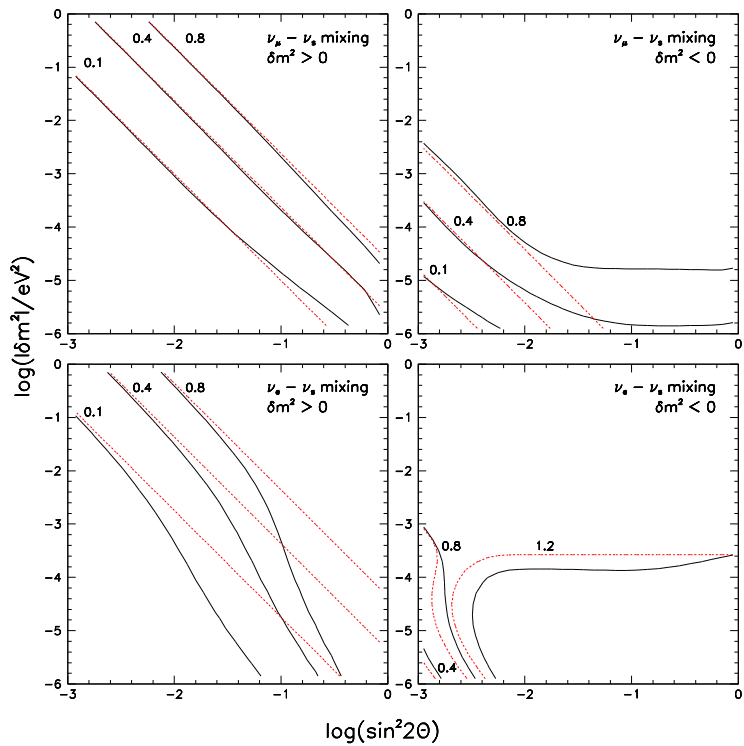} 
\end{center} 
\caption[]{BBN bounds on active-sterile neutrino mixing. See \cite{dolgovil}
for details} 
\label{f2} 
\end{figure}

 The results of \cite{dolgovil} are shown in Fig.~\ref{f2}. The effect on 
BBN is expressed in term of variation of the effective number of neutrinos 
$\Delta N_{\nu}$. The upper panels are 
for the case of $\num-\nus$ (or $\nut-\nus$) mixing, while the lower  
panels refer to the case of $\nue-\nus$ mixing. 
The obtained results clearly depend on the sign of the mass differences.  
For positive\footnote{In the notations of \cite{dolgovil},  
$\delta m^2$ is positive if sterile neutrino is heavier than  
active neutrino, in the limit $\theta\rightarrow 0$.} mass differences, 
$\delta m^2>0$ (left panels), sterile neutrino production occurs through 
non-resonant transitions. For $\delta m^2<0$ (right panels), one has  
instead resonant active-sterile transitions which result in much stronger  
bounds on the neutrino oscillation parameters. 
The solid lines in fig.\ref{f2} correspond to numerical results, while the  
red dotted lines corresponds to analytic approximate results.
It is evident that an observational bound on extra neutrinos much better
than unity, say $\delta N_{\nu}<0.3$, could give very restrictive limits on
active-sterile neutrino mixing. Unfortunately, the present observational 
bound $\delta N_{\nu}\le 1.0$ is not accurate enough to put relevant 
constraints.

\subsection{Three active and one sterile neutrinos} 
 
It is practically established now that all active neutrinos are mixed with 
parameters given by the Large Mixing Angle solution to solar neutrino 
problem ($\delta m^2_{\rm sol}=7.3\cdot 10^{-5}\;{\rm eV}^2 
$ and $\tan^{2}\theta_{\rm sol}=0.4$) and by atmospheric neutrino  
data ($\delta m^2_{\rm atmo}=2.5\cdot 10^{-3}\;{\rm eV}^2$ and  
$\tan^{2}\theta_{\rm atmo}\approx 1$).  
Existence of fast transitions between $\nue$, $\num$, and $\nut$  
may noticeably change BBN bounds on mixing with sterile neutrinos,  
expecially for small values of mass difference.  
In particular, due to oscillations between active neutrinos, 
a deficit of $\num$ or $\nut$ would be efficiently  
transformed into a deficit of $\nue$, leading to stronger bounds  
on active-sterile mixing.  
The effects of mixing between active neutrinos  
on the BBN bounds on a possible active-sterile admixture  
has been investigated in detail in \cite{dolgovil}. 
 
\section{Conclusion} 
Comparison of BBN theoretical results 
with observational data is not straightforward because the data 
are subject to poorly known evolutionary effects and systematic errors. 
Still, even with these uncertainties, BBN permits to constraint important 
cosmological parameters, like e.g. the baryon density $\Omega_{\rm B}h^2$,  
the effective number neutrino families $N_{\nu}$, the neutrino degeneracy  
parameters $\xi_{a}$ etc. 
The present bound on the number of extra neutrinos species $\delta N_{\nu}$  
is about unity and is not accurate enough to put relevant constraints on  
active-sterile neutrino mixing.  
If this limit could be reduced in the next future, 
say to $\delta N_{\nu}<0.3$, very restrictive limits on active-sterile 
admixture could be obtained.

%
 
\end{document}